\documentclass[]{aa}

\usepackage{graphicx}
\usepackage{txfonts}
\usepackage{lipsum}
\usepackage{subcaption}         
\usepackage{lscape}            
\usepackage{placeins}           
\usepackage[colorlinks,citecolor=blue]{hyperref}
\hypersetup{
    colorlinks = true,
    linkcolor = blue,
    anchorcolor = blue,
    citecolor = blue,
    filecolor = blue,
    urlcolor = blue
    }
\usepackage{booktabs}
\usepackage{mathrsfs}
\usepackage{tikz}
\usepackage{booktabs}
\usetikzlibrary{calc}
\usepackage{mathtools}
\usepackage{siunitx}
\usepackage{svg}
\usepackage{ulem}

\begin{document} 

   \title{Discovery of the $\lambda$~Bootis phenomenon in an early B-type star}

   \author{J. Olbermann\inst{1} \email{Janick.Olbermann@student.uibk.ac.at}
          \and
          P. Aschenbrenner \inst{1} \email{Patrick.Aschenbrenner@student.uibk.ac.at}
          \and
          N. Przybilla\inst{1} \corrauth{norbert.przybilla@uibk.ac.at}
          }

   \institute{Universit\"at Innsbruck, Institut f\"ur Astro- und Teilchenphysik, Technikerstr. 25/8, 6020 Innsbruck, Austria
             }

   \date{Received ; accepted }

  \abstract
   {Chemically peculiar A-type stars of the $\lambda$~Boo class are characterised by depletions of refractory elements and normal volatile species. Their formation channel(s) have not yet been fully established, but they provide insight into star-disc-planet interactions during star formation, star-planet interactions in later system evolution, or interactions between stars and the interstellar medium.}
   {We report the first detection of the $\lambda$ Boo phenomenon in a massive early B-type star, HD~37356, a member of Ori OB1c.}
   {We analysed a high-quality spectrum using a hybrid non-LTE model-atmosphere technique. We determined atmospheric parameters and elemental abundances using neural networks to emulate synthetic spectra, coupled with a Markov chain Monte Carlo algorithm. We also derived the fundamental parameters of the star using \textit{Gaia} data and stellar evolution models.}
   {HD~37356 closely resembles the Morgan-Keenan standard $\gamma$~Peg, closely agreeing in spectral signatures except for the lines of refractory elements, which are systematically weaker. The quantitative analysis yields similar stellar parameters. The abundances of volatile elements are compatible with cosmic standard abundances, while refractory species are depleted by 0.25 to 0.45\,dex. In particular, the lower iron abundance explains the different pulsational behaviour of HD~37356 compared to $\gamma$~Peg. Of the different formation channels for $\lambda$~Boo stars, only the accretion of gas depleted in refractory elements from the circumstellar disc during star formation appears to be feasible in this case. This scenario requires HD~37356 to be a true slow rotator in order to prevent the abundance pattern from being erased by internal mixing processes such as meridional circulation.} 
   {}


   \maketitle
   \nolinenumbers

\section{Introduction}
The $\lambda$~Boo stars are the third most common chemically peculiar (CP) A-type stars after the slow-rotating Am and magnetic Ap stars and are found between spectral types B9 and F3 \citep{Corbally18}. They are characterised by a deficiency of heavy elements in their spectra, while the abundances of light elements carbon, nitrogen, oxygen, and sulphur remain close to solar \citep{Heiter02}. This resembles the general pattern of metal depletion found in the interstellar medium (ISM), with refractory elements locked in dust grains and only volatile elements remaining in the gas phase \citep{VeLa90}. The $\lambda$ Boo phenomenon affects A-type stars of all ages, from star formation \citep[see examples in OB associations;][]{GrCo93} to the terminal-age main sequence. However, the cause of the chemical peculiarity remains unclear, and multiple channels can produce the spectral signature of $\lambda$~Boo stars \citep{MuPa17}.

Based on observed abundance patterns, \citet{VeLa90} suggested that $\lambda$~Boo stars result from the separation of circumstellar, or interstellar, gas from grains, and subsequent accretion of the gas by the star \citep[see also][]{Watersetal92}. The bulk of the interstellar grains would then comprise a circumstellar cloud or disc that would be detectable by its IR radiation. Indeed, the incidence of bright IR excesses around $\lambda$~Boo stars is higher than in normal A-type stars, which is interpreted as originating from debris discs \citep{Draperetal16}.  \citet{Kamaetal15} proposed a mechanism in which a Jupiter-type planet blocks the flow of dust from a protoplanetary disc to the stellar surface late in its formation. Later in the life of A-type stars, accretion of material from the winds of hot Jupiters could also lead to the $\lambda$~Boo abundance pattern \citep{Jura15}; radiation pressure on gas-phase ions might selectively allow the accretion of light elements with relatively few spectral lines such as carbon, nitrogen, and oxygen, while the inflow of elements such as iron (with its rich spectrum) may be suppressed -- the presence of dust would not be required in that case.  \citet{KaPa02} suggested the interaction between a star and a diffuse interstellar cloud  as the origin of the $\lambda$~Boo abundance pattern. Once the interaction ends, the resulting abundance pattern would remain visible for only a short timescale of $\sim$10$^6$\,yr, and would be gradually diminished by meridional circulation in intermediate- to fast-rotating $\lambda$~Boo stars \citep{Turcotte02}, explaining the rarity of this kind of star. 

The opportunity to study star-disc-planet interactions during star formation, star-planet interactions in later system evolution, or star-ISM interactions makes $\lambda$~Boo stars interesting in the astrophysical context beyond their immediate peculiarity. Progress in understanding complex problems often comes from a perspective that is independent of the immediate context. Here, we report the serendipitous discovery of a $\lambda$~Boo abundance pattern in a massive early B-type star that, compared to A-type stars, has a substantial stellar wind. The paper is structured as follows: We introduce the observational material in Sect.~\ref{sec:obs} and the analysis methodology in Sect.~\ref{section:analysis}.  Section~\ref{section:results} presents the results, and we draw our conclusions in Sect.~\ref{sec:conclusions}.

\begin{figure*}[ht]
\centering
\includegraphics[width=.94\linewidth]{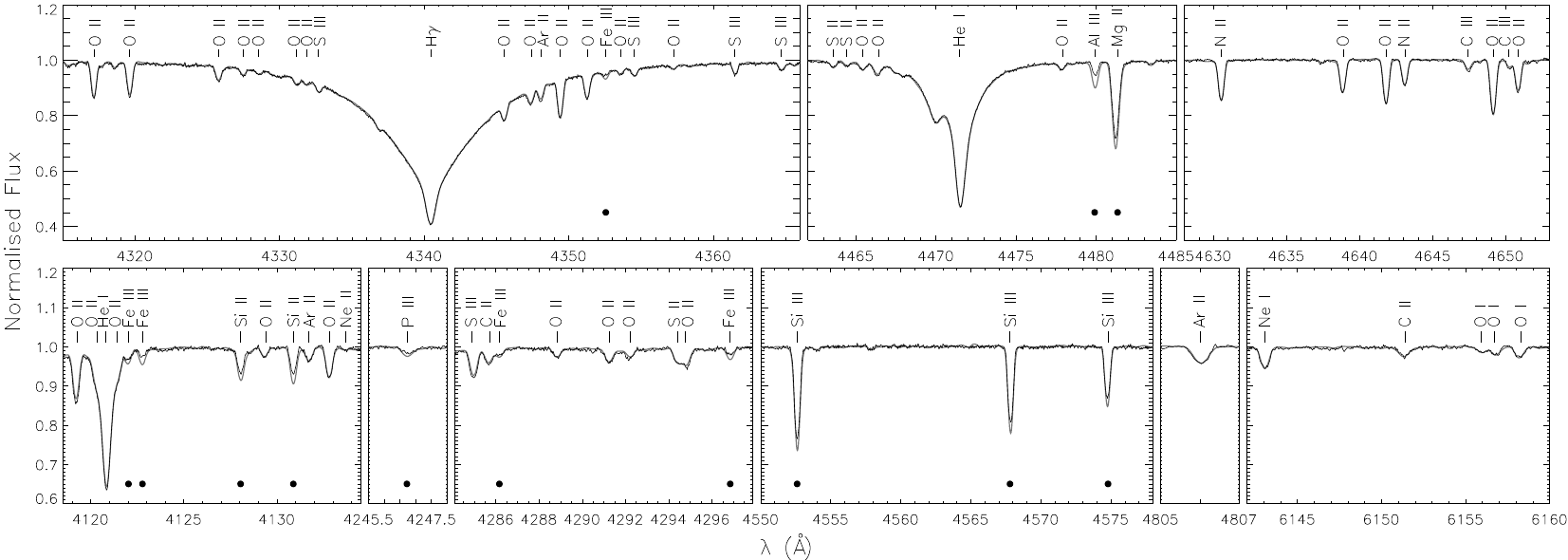}
\caption{Comparison of strategic regions of the observed spectra of HD~37356 (black line) and $\gamma$~Peg (grey line), covering all analysed chemical species. The spectrum of $\gamma$~Peg was artificially broadened to match that of HD~37356. Both spectra were shifted to the laboratory rest frame. Line identifications are given, and lines of refractory elements are marked with dots.}
\label{fig:comparison}
\end{figure*}

\section{Observational data}\label{sec:obs}
Our study is based on a spectrum of \object{HD 37356} observed on the night of 8 November 2011 (programme ID 088.D-0064, P.I.~A.~Irrgang) with the Fibrefed Extended~Range Optical Spectrograph \citep[{FEROS};][]{Kauferetal99} mounted on the Max-Planck-Gesellschaft/European Southern Observatory (ESO) 2.2\,m telescope at La Silla in Chile. We downloaded pipeline-reduced Phase~3 data from the ESO science portal\footnote{\rule{-0.25mm}{0mm}\url{https://archive.eso.org/scienceportal/home}}. It covers a useful wavelength range from about 3700 to 9200\,{\AA} at a resolving power of $R$\,=\,$\lambda / \Delta \lambda$\,$\approx$\,48\,000 and a signal-to-noise ratio of $S/N$\,$\approx$\,400 per pixel, measured around 5000\,{\AA}. We normalised the spectrum by fitting a spline function through carefully selected continuum points. For comparison, we adopted a spectrum of $\gamma$~Peg (\mbox{\object{HD 886}}) as observed with the Fibre Optics Cassegrain Echelle Spectrograph \citep[FOCES;][]{Pfeifferetal98} on the Calar Alto 2.2m telescope in Spain, with $R$\,$\approx$\,40\,000 and $S/N$$\approx$\,500 \citep[for further details, see][]{NiPr12}.

Moreover, for the construction of the stellar energy distribution (SED), we used low-dispersion, large-aperture spectrophotometry for HD~37356 (IDs SWP40064 and LWP19162) obtained with the International Ultraviolet Explorer, IUE\footnote{\url{https://archive.stsci.edu/iue/}}. In the optical we also considered a low-resolution spectrum from \textit{Gaia} Data Release~3 \citep[DR3;][]{Gaia2016,Gaia23} and Johnson $UBV$ magnitudes \citep{Mermilliod97}. In the IR, we adopted $JHK$ magnitudes from the Two Micron All Sky Survey \citep[2MASS;][]{Cutrietal03} and AllWISE photometry \citep{ALLWISE_vizier}, based on data from the Wide-Field Infrared Survey Explorer (WISE) mission. 
Finally, we adopted time-series photometry from the Transiting Exoplanet Survey Satellite (TESS), extracted via the Mikulski Archive for Space Telescopes (MAST)\footnote{\url{https://mast.stsci.edu/portal/Mashup/Clients/Mast/Portal.html}}.

\section{Analysis}\label{section:analysis}
We employed hybrid non-local thermodynamic equilibrium (non-LTE) model atmospheres computed with the codes {\sc Atlas} \citep{Kurucz93,Kurucz96}, {\sc Detail} \citep{Giddings81}, and {\sc Surface} \citep{BuGi85}. The general modelling approach was described by \citet{NiPr07,NiPr12}, with further refinements described by \citet{Aschenbrenneretal23}. For the analysis we used {\sc Saturn} \citep[Spectral Analysis Tool Using Restricted Neural networks;][]{Aschenbrenneretal26}, which couples emulation of synthetic spectra using previously trained neural networks with a Markov chain Monte Carlo algorithm to determine atmospheric parameters -- effective temperature $T_\mathrm{eff}$, (logarithmic) surface gravity $\log g$ (in cgs units), helium abundance (by number) $y$, microturbulence $\xi$, projected rotational velocity $\varv \sin i$, (radial-tangential) macroturbulence $\zeta$, and radial stellar velocity $\varv_\mathrm{rad}$ --, and the abundances for eleven chemical species via spectrum fitting, including a realistic uncertainty determination.

To determine the fundamental parameters, we employed stellar evolution tracks of \cite{Ekstroemetal12} for non-rotating stars at metallicity $Z$\,=\,0.014 \citep[assuming the $\lambda$~Boo phenomenon affect only the surface, whereas the bulk composition follows the present-day cosmic abundance standard, CAS;][]{NiPr12}. We determined the evolutionary mass $M$ by interpolating the tracks in the spectroscopic Hertzsprung-Russell diagram \citep[sHRD;][]{LaKu14}. We determined the evolutionary age $\tau$ by interpolating isochrones based on the Ekström et al. tracks in the Hertzsprung-Russell diagram (HRD), and we determined the fractional main-sequence lifetime $\tau/\tau_\mathrm{MS}$ by comparing this evolutionary age with the terminal main sequence age of an appropriate (interpolated) evolutionary track. Based on the evolutionary mass and $\log g$, we calculated the stellar radius $R$ and determined the luminosity using the Stefan-Boltzmann law.

Finally, using the reddening law of \cite{fitzpatrick99}, we constrained interstellar reddening $E\left(B-V\right)$ and the ratio of total-to-selective extinction $R_V$ by fitting a model SED, computed using the {\sc Atlas9} code, to the photometric and spectrophotometric data described in the previous section. We calculated the absolute bolometric magnitude $M_{\mathrm{bol}}$ from the luminosity ($M_{\mathrm{bol}}$\,=\,$-2.5\cdot\log\left( L/L_\sun \right)+M_{\mathrm{bol},\sun}$, where $M_{\mathrm{bol},\sun}$\,=\,4.74\,mag) and determined the absolute visual magnitude $M_V^0$ using the bolometric correction from our model SED. In addition, we derived the spectroscopic distance $d_\mathrm{spec}$ using Eq. (2) of \cite{Aschenbrenneretal23}.

\begin{table}
\caption{Atmospheric and stellar parameters of HD~37356.}
\vspace{-0.6cm}
\label{tab:parameters}
{\small
\setlength{\tabcolsep}{1mm}
\begin{center}    
\begin{tabular}{ll@{\hspace{3.5mm}}ll}        
\hline\hline
\multicolumn{4}{l}{General information:} \\
Sp. Type & B2\,IV\tablefootmark{a}    &   $d_\mathrm{Gaia}$\,(pc)\tablefootmark{c} & 450$^{+13}_{-10}$\\
OB Assoc. & Ori OB1c\tablefootmark{b} &  $d_\mathrm{spec}$\,(pc)  & 380$^{+15}_{-15}$ \\
$\varv_\mathrm{rad}$ & 29.3$\pm$0.5\,km\,s$^{-1}$ \\[1.5mm]
\multicolumn{4}{l}{Atmospheric parameters:} \\
$T_{\mathrm{eff}}$\,(K)    &  22039$\pm$131    &  $\xi$\,(km\,s$^{-1}$)           & 1.6$\pm$0.8\\
$\log g$\,(cgs)            & 3.96$\pm$0.02     &  $\varv  \sin i$\,(km\,s$^{-1}$) &  18.6$\pm$0.6\\
$y$\,(by number)           & 0.096$\pm$0.002    &  $\zeta$\,(km\,s$^{-1}$)          &  3.8$\pm$2.4\\[1.5mm]
\multicolumn{4}{l}{Elemental abundances:} \\
             & $\log$\,($X$/H)\,+\,12 & \multicolumn{2}{c}{CAS\tablefootmark{d}}\\ \cline{2-4}
C            & 8.43$\pm$0.04\,(10)     & \multicolumn{2}{c}{8.33$\pm$0.04~~}\\  
N            & 7.81$\pm$0.02\,(31)     & \multicolumn{2}{c}{7.79$\pm$0.04~~}\\
O            & 8.79$\pm$0.08\,(59)     & \multicolumn{2}{c}{8.76$\pm$0.05~~}\\
Ne           & 8.05$\pm$0.03\,(12)     & \multicolumn{2}{c}{8.09$\pm$0.05~~}\\
Mg           & 7.31$\pm$0.04\,(6)     & \multicolumn{2}{c}{7.56$\pm$0.05~~}\\
Al           & 5.88$\pm$0.06\,(4)     & \multicolumn{2}{c}{6.30$\pm$0.07$^*$}\\
Si           & 7.25$\pm$0.04\,(11)     & \multicolumn{2}{c}{7.50$\pm$0.05~~}\\
P            & 4.94$\pm$0.07\,(3)     & \multicolumn{2}{c}{5.36$\pm$0.14~~}\\
S            & 7.08$\pm$0.09\,(4)     & \multicolumn{2}{c}{7.14$\pm$0.06$^*$}\\
Ar           & 6.58$\pm$0.05\,(9)     & \multicolumn{2}{c}{6.57$\pm$0.02$^*$}\\
Fe           & 7.07$\pm$0.06\,(15)     & \multicolumn{2}{c}{7.52$\pm$0.03~~}\\[1.5mm]
\multicolumn{4}{l}{Photometric data:} \\
$V$\,(mag)\tablefootmark{e}   & ~~\,6.188$\pm$0.015 &$R_V$ & 3.63$\pm$0.09\\
$B-V$\,(mag)\tablefootmark{e} & $-$0.041$\pm$0.003  & $M_V^0$\,(mag)     & $-$2.44$^{+0.07}_{-0.07}$\\
$E\left(B-V\right)$\,(mag) & ~~0.209\,$\pm$0.004 &  $M_{\mathrm{bol}}$\,(mag) & $-$4.63$^{+0.05}_{-0.05}$\\[1.5mm]
\multicolumn{4}{l}{Fundamental stellar parameters:} \\
$M/M_{\odot}$   & 8.77$^{+0.07}_{-0.06}$ & $\tau$\,(Myr)           & 20.2$\pm$0.6\\
$R/R_{\odot}$   & 5.15$^{+0.13}_{-0.13}$ & $\tau/\tau_\mathrm{MS}$ & 0.73$^{+0.02}_{-0.02}$\\
$\log L/L_\sun$ & 3.75$^{+0.02}_{-0.02}$ & \\[0.3mm]
\hline
\end{tabular}
\end{center}}

\tablefoot{Uncertainties are standard deviations. $^*$ Preliminary values. Systematic errors of the abundances due to uncertainties of the atmospheric parameters, atomic data, and continuum normalisation amount to $\sim$0.1\,dex \citep{Przybillaetal00,Przybillaetal01a,Przybillaetal01b}.
\tablefoottext{a}{This work, due to resemblance to the B2\,IV MK standard $\gamma$~Peg} 
\tablefoottext{b}{\citet{Brownetal94}}
\tablefoottext{c}{\citet[photogeometric\,distance]{Bailer-Jones_etal_2021}} 
\tablefoottext{d}{\citet{NiPr12}, \citet{Przybillaetal13}, \citet{Aschenbrenneretal25}}
\tablefoottext{e}{\cite{Mermilliod97}.}
}
\end{table}

\begin{figure*}[ht]
\sidecaption
\includegraphics[width=12cm]{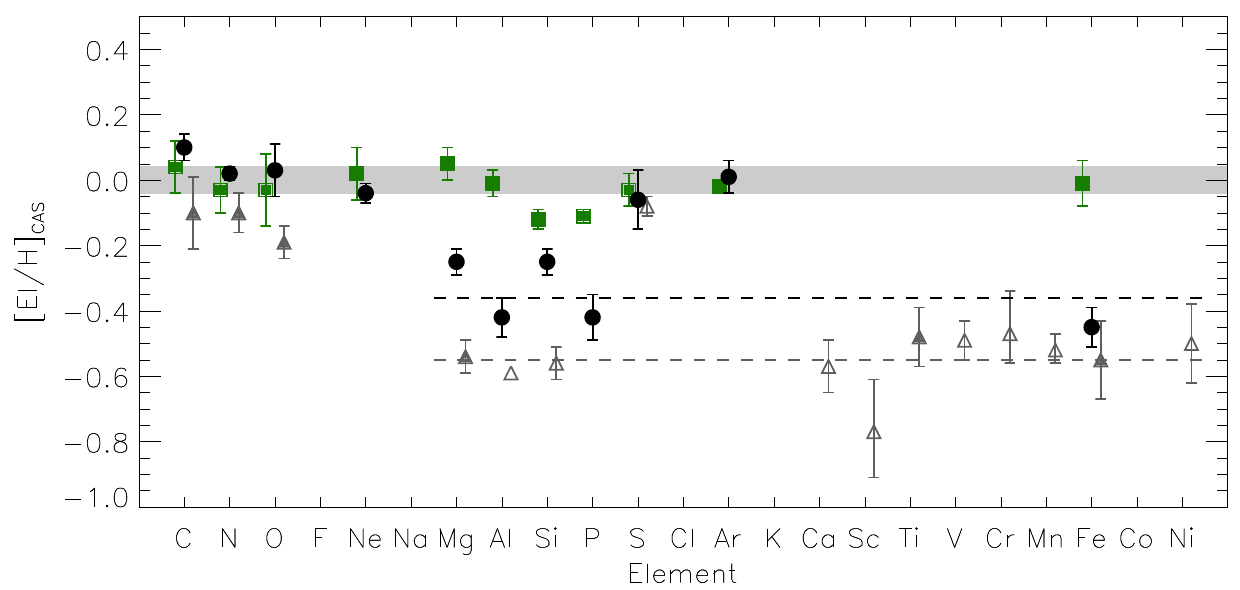}
\caption{Elemental abundances relative to the cosmic abundance standard \citep[CAS;][grey band]{NiPr12,Przybillaetal13,Aschenbrenneretal25}, as derived from OB-type stars in the solar neighbourhood, and supplemented, where unavailable, by solar meteoritic data \citep{Asplundetal21}. Data for HD~37356 (black dots), $\gamma$~Peg \citep[green boxes;][and this work]{NiPr12} and the mild $\lambda$~Boo star Vega \citep[grey triangles]{Przybilla02} are shown. Filled symbols mark non-LTE and open symbols LTE abundances. Error bars denote 1$\sigma$ uncertainties. The dashed lines indicate the average depletions of refractory elements in  HD~37356 (black) and Vega (grey).}
\label{fig:abundances}
\end{figure*}

\section{Results}\label{section:results}
The similarity between the spectra of HD~37356 and $\gamma$~Peg over wide parts of the wavelength range is remarkable. Figure~\ref{fig:comparison} shows a comparison of strategic spectral regions of both stars; for this,  we artificially broadened the spectrum of $\gamma$~Peg to match the broadening of HD~37356. Lines of volatile species H, He, C, N, O, Ne, S, and Ar match almost perfectly, with the Balmer lines indicating very similar surface gravity, and the matching ionisation equilibria of \ion{C}{ii/iii}, \ion{O}{i/ii}, \ion{Ne}{i/ii}, and \ion{S}{ii/iii} implying very similar $T_\mathrm{eff}$. Moreover, nearly identical \ion{He}{i} lines indicate similar helium abundance, and the overall match of weak and strong lines of the volatile species yields a similar microturbulent velocity. However, the lines of the refractory elements Mg, Al, Si, P, and Fe are systematically weaker in HD~37356, which requires lower abundances. Hence, the signature of a $\lambda$~Boo abundance pattern is evident on pure observational grounds. However, we note that the differences are subtle enough to be overlooked at lower spectral resolution and $S/N$, or at higher $\varv \sin i$. Moreover, since the effective temperature differs substantially from that of A-type $\lambda$~Boo stars, classical line ratios used for empirical classification as $\lambda$~Boo peculiar in the optical \citep{Paunzenetal14,Chengetal17} and in the UV \citep{SoPa98,SoPa99} cannot be applied. This is because lines from species such as \ion{C}{i} or \ion{Fe}{i} are absent from the spectrum, and the \ion{Ca}{ii} H+K lines are of interstellar origin.

We derived the exact depletion of the refractory elements by quantitative analysis. Table~\ref{tab:parameters} summarises the results for all constrained parameters of HD~37356, which also provides values for the CAS abundances; Fig.~\ref{fig:specfits} shows the high-quality fit of the model to the observed spectrum. The atmospheric parameters, abundances of volatile species, and fundamental parameters of HD~37356 agree with those of $\gamma$~Peg \citep{NiPr12,NiPr14} within the uncertainties. The abundances of the refractory elements are depleted by $\sim$0.25 to 0.45\,dex. This is visualised in Fig.~\ref{fig:abundances}, which also compares the abundance pattern with that of the fast-rotating, mild $\lambda$~Boo star \object{Vega} \citep{Przybilla02}. Non-LTE abundances are not available for all species. The depletion of refractory elements is less in HD~37356 than in Vega, and it is less uniform. For neon and argon, the hotter star provides access to highly volatile noble gas species, which are, as expected, not depleted.

\begin{figure}[ht]
\centering
\includegraphics[width=.9\linewidth]{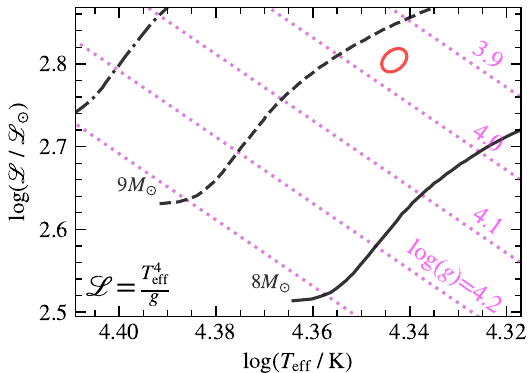}
\includegraphics[width=.9\linewidth]{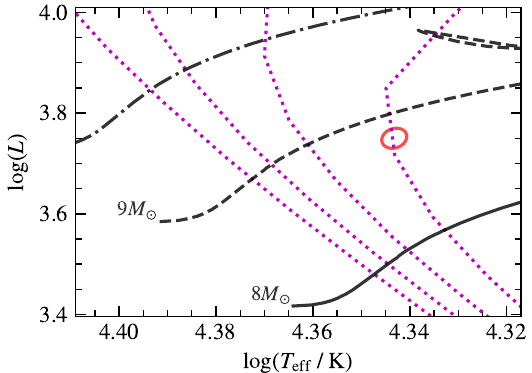}
\caption{Position of HD~37356 (red ellipse) in the sHRD and HRD. Evolutionary tracks from \cite{Ekstroemetal12} for non-rotating stars at $Z$\,=\,0.014 with $8$, $9$, and $10$ $M_{\sun}$ are shown as solid, dashed, and dashed-dotted black lines, respectively. \textit{Top}: Position of the star in the sHRD. Lines of constant $\log g$ are indicated by dotted magenta lines. \textit{Bottom}: Position of the star in the HRD. Isochrones from \cite{Ekstroemetal12} for $\log\tau_{\mathrm{evol}}(\si{yr})$\,=\,7.0, 7.1, 7.2, and 7.3 are overlaid (dotted magenta lines).}
\label{fig:HRD}
\end{figure}

In the past, studies of early B-type stars in the solar neighbourhood \citep{Kilian92,Kilian94} and in the Orion OB1 association \citep{CuLa94} considered them to be metal-poor compared to the Sun and chemically highly inhomogeneous. This changed once photometric techniques for parameter determinations were abandoned and self-consistent spectroscopic analysis techniques were employed in combination with atomic data from ab initio calculations on a large scale \citep{Przybillaetal08b,NiPr12}, indicating chemical homogeneity in the solar neighbourhood and abundance values close to solar values (i.e. CAS values). For the Orion association, \citet{Simon-Diaz10} deduced chemical homogeneity and elemental abundances compatible with CAS values using 13 targets in the four Ori OB1 sub-groups, and \citet{NiSi11} confirmed and extended this using two independent non-LTE model atmosphere techniques. This also applies to the components of a multiple system in Orion \citep{AschPr24}. Compared to these and many other OB stars in the solar neighbourhood, the abundance pattern of HD~37356 reported here is unusual, even considering systematic abundance uncertainties of $\sim$0.1\,dex. We discuss the results of the present analysis for HD~37356 in the context of previous studies in Appendix~\ref{appendixA}.

Figure~\ref{fig:HRD} shows the position of HD~37356 in the sHRD and in a classical HRD. The position in both plots agrees with the stellar evolution tracks when the spectroscopic distance is adopted. Use of the \textit{Gaia} distance of 450\,pc would result in no match, corresponding to $\log L/L_\odot$\,=\,3.9. However, that distance would put HD~37356 beyond the Orion Trapezium cluster at about 390\,pc \citep{Dzibetal26}, that is, inside the Orion giant molecular cloud, which is clearly not the case. Therefore, despite a `good' renormalised unit weight error (RUWE) of 1.079, it is likely that systematic effects affect the \textit{Gaia} DR3 parallax of this bright star, affecting both the absolute value and the uncertainty \citep[see e.g.][]{MaizApellaniz22}. 

\begin{figure}[ht]
\centering
\includegraphics[width=.93\linewidth]{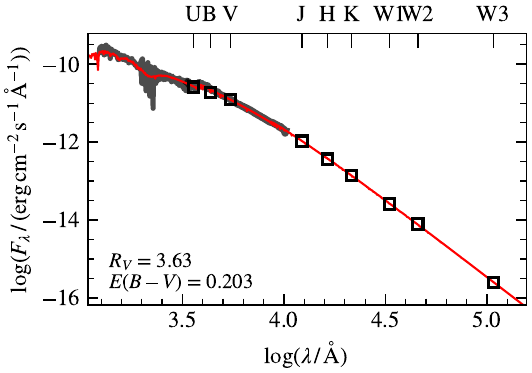}
\caption{Spectral energy distribution fit of the reddened {\sc Atlas9} model flux (red)
to photometric measurements (squares), and to observed IUE spectrophotometry and Gaia XP spectra (grey).}
\label{fig:SED}
\end{figure}

\begin{figure}[ht]
\centering
\includegraphics[width=.99\linewidth]{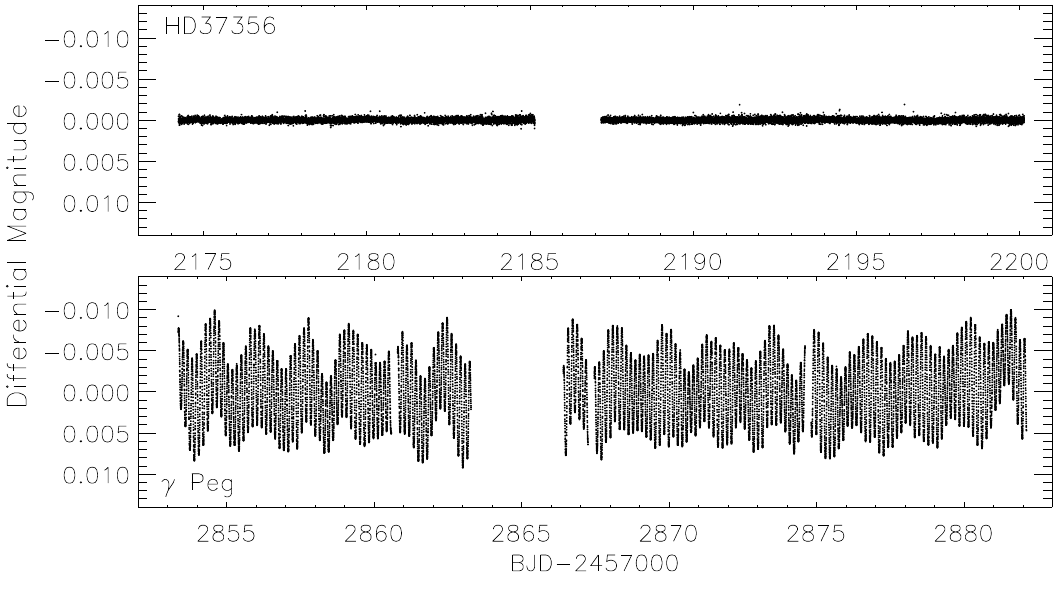}
\caption{Differential magnitudes from TESS observations of HD~37356 in Sector~32 (\textit{top}) and of $\gamma$~Peg from Sector~57 (\textit{bottom}). The abscissa shows the barycentric Julian date. The median values of the measurements are set to zero.}
\label{fig:variability}
\end{figure}

Figure~\ref{fig:SED} shows a fit of our model flux to the observed SED of HD~37356, confirming that our solution reproduces not only the optical spectrum, but also the global energy output of the star. The reddening law deviates slightly from standard, but this is not unexpected for an object close to the Orion molecular cloud. We did not include the AllWISE W4 data point as it is affected by background dust emission, which also complicates the detection of a potential debris disc around the star via excess radiation in the far-IR.

Figure~\ref{fig:variability} shows the variability of HD~37356 in the TESS time-series photometry compared to that of $\gamma$~Peg. The light curves of both targets are not affected by other stars within the large TESS pixels owing to their brightness. While $\gamma$~Peg shows pronounced variability as a hybrid pulsator showing characteristics of both $\beta$~Cep and slowly pulsating B-star (SPB) pulsations \citep{Chapellieretal06,Handleretal09}, HD~37356 is stable at a level of less than about 0.6\,mmag (3$\sigma$-value). The $\kappa$-mechanism driving both p-mode and g-mode $\beta$~Cep and SPB pulsations operates at the iron opacity bump at temperatures near 200\,kK. In the two stars that are otherwise very similar, the significantly reduced iron abundance in HD~37356 may explain the difference in pulsational behaviour. This also indicates that the $\lambda$~Boo abundance pattern extends at least to the outer envelope layers where the iron bump operates.

Finally, we discuss the potential binarity of HD~37356. For decades, a constant radial velocity $\varv_\mathrm{rad}$\,=\,29.1$\pm$2.6\,km\,s$^{-1}$ was reported \citep{PlPe31,Gontcharov06}, and the small RUWE value of the \textit{Gaia} parallax disfavours binarity. However,  \citet{Cottaaretal15} published pipeline-determined radial velocities from 14 near-IR spectra from the Sloan Digital Sky Survey III (SDSS-III) Apache Point Observatory Galactic Evolution Experiment (APOGEE), suggesting changes in $\varv_\mathrm{rad}$ of more than 130\,km\,s$^{-1}$ (with typical uncertainties of  a few km\,s$^{-1}$) within a time span of about 50~days, and almost 100\,km\,s$^{-1}$ within one day. This appears unplausible, and, indeed, a later study of the same data using a different analysis pipeline derived a smaller range of $\varv_\mathrm{rad}$ between 11 and 66\,km\,s$^{-1}$ (with uncertainties typically between 10 and 20\,km\,s$^{-1}$, but up to 117\,km\,s$^{-1}$). Based on this, the agreement of our $\varv_\mathrm{rad} $ measurement (see Table~\ref{tab:parameters}) with the classical value and based on the quality of our SED fit, we see no evidence of a companion to HD~37356. We note, however, that analysis of an object with a second continuum contribution as a single star can lead to underestimated abundances. The low abundances of refractory elements in HD~37356 cannot be explained that way, as volatile species would experience the same effect. However, we encourage a careful study of the radial velocity of HD~37356 to clarify the binarity issue and  search for a potential substellar companion, as well as an investigation using high-spatial-resolution techniques. Interferometric observations have recently revealed that $\gamma$~Peg is a binary, with the secondary contributing to the light in the $H$ band by 3.6$^{+0.1}_{-0.4}$\% \citep{Frostetal25}. The secondary must therefore be a significantly cooler star, implying negligible second light in the optical spectrum.

\section{Conclusions}\label{sec:conclusions}
Having discovered the presence of a $\lambda$~Boo abundance pattern in the atmosphere of HD~37356, its origin remains to be addressed. Scenarios involving the accretion of volatile-depleted material from the passage of a diffuse interstellar cloud or from the wind of a hot Jupiter are unlikely for an early B-type star. A-type ($\lambda$~Boo) stars have fractionated metal winds with a mass-loss rate $\dot{M}$ that falls below 10$^{-16}$\,$M_\odot$\,yr$^{-1}$ \citep{Babel95}, which allows the accretion of gas for sufficiently large accretion rates. \citet{KaPa02} found that 10$^{-13}$\,$M_\odot$\,yr$^{-1}$ at a relative velocity of 10\,km\,s$^{-1}$  between an ISM cloud and a star was sufficient. A B2\,IV star such as HD~37356 is expected to have an ionised homogeneous wind (i.e. hydrogen and helium are Coulomb-coupled to the wind-driving metals) with $\dot{M}$\,$\simeq$\,10$^{-10}$\,$M_\odot$\,yr$^{-1}$ \citep{Krticka14}, which is too strong for accretion to occur.

Accretion from a circumstellar disc, with or without a Jovian planet or a brown dwarf, remains a plausible scenario for HD~37356 and links its $\lambda$ Boo abundance pattern to the late stages of stellar formation. Dust trapped in pressure maxima of circumstellar discs has been described theoretically \citep[e.g.][]{Pinellaetal12} and has also been observed \citep[e.g.][]{Krausetal17}.

The challenge here is to preserve the abundance signature over a lifetime of $\sim$20\,Myr. One issue is the stellar wind, which has removed on the order of 0.002\,$M_\odot$ of material from the star; this provides a lower limit to the mass of refractory-poor material that must be accreted initially. The other issue is internal mixing processes, such as meridional circulation, which can dilute surface signatures over time. HD~37356 appears to be a true slow rotator, since the  C, N, and O abundances remain at the starting point of the nuclear path expected for mixing with CNO-processed material \citep{Przybillaetal10,Maederetal14}. The $\varv \sin i$ of HD~37356 is at the very low end of the observed velocity distribution of OB-type stars \citep[e.g.][]{deBurgosetal24}, making the star special. 

As accretion of matter from a circumstellar disc also adds angular momentum, HD~37356 must have experienced efficient removal of angular momentum during its formation. It is beyond the scope of the present paper to investigate scenarios for the evolution of the star in detail, but we outline potential options for angular-momentum removal. These include transfer to the orbital motion of a substellar companion, removal via a protostellar jet \citep{Olivaetal26}, or coupling between a potential stellar magnetic field (testable via spectropolarimetry) and the disc, which would lead to efficient spin-down. However, the latter scenario is unlikely, since HD~37356 lies in a region of the HRD where strongly magnetic stars are expected to exhibit He-strong characteristics \citep[e.g.][]{HuGr99,Przybillaetal16}.

A final issue arising from this result is whether other early B-type stars show the $\lambda$~Boo phenomenon. HD~37356 may lie towards the hot end of the temperature range where the $\lambda$~Boo phenomenon occurs. Hotter stars emit harder UV radiation during formation, which efficiently photoevaporates dust. A possible indicator is the abundances of magnesium and silicon (see Fig.~\ref{fig:comparison}), which implies that part of the magnesium-containing silicate dust component less stable than pyroxene (MgSiO$_3$) and olivine (Mg$_2$SiO$_4$) may have entered the gas phase, leading to reduced depletion. In contrast, corundum (Al$_2$O$_3$) is one of the most thermally stable compounds in circumstellar discs, so little aluminium would be released into the gas phase; metallic iron behaves similarly. To fully verify the viability of the observed abundance pattern, we would need chemodynamical calculations for a circumstellar disc that locate dust traps relative to the sublimation boundaries of different dust components; this is beyond the scope of this paper. 

At cooler effective temperatures, in the spectral gap towards the classical $\lambda$~Boo stars, $\lambda$ Boo signatures imprinted during star formation may initially also be present in parallel to HD~37356. However, they must subsequently compete with atomic diffusion, which affects all slowly rotating B-type CP stars, and the $\lambda$~Boo signature is erased.

\section*{Data availability}
The observational material employed in this work is accessible via the ESO science portal, the Mikulski Archive for Space Telescopes, and the VizieR Catalogue Service. The FOCES spectrum of $\gamma$~Peg is available from the authors upon reasonable request.

\begin{acknowledgements}
The authors thank E.~Vorobiev for fruitful discussion and the referee E.~Paunzen for valuable suggestions. Based on observations collected at the European Southern Observatory under ESO programme 088.D-0064(A). Based on observations collected at the Centro Astron\'omico Hispano Alem\'an at Calar Alto (CAHA), operated jointly by the Max-Planck Institut f\"ur Astronomie and the Instituto de Astrof\'isica de Andaluc\'aa (CSIC), proposal H2005-2.2-016.
This work has made use of data from the European Space Agency (ESA) mission {\it Gaia} (\url{https://www.cosmos.esa.int/gaia}), processed by the {\it Gaia} Data Processing and Analysis Consortium (DPAC, \url{https://www.cosmos.esa.int/web/gaia/dpac/consortium}). Funding for the DPAC has been provided by national institutions, in particular the institutions participating in the {\it Gaia} Multilateral Agreement.
This paper includes data collected by the TESS mission. Funding for the TESS mission is provided by the NASA's Science Mission Directorate.
\end{acknowledgements}

\clearpage

\typeout{}
\bibliographystyle{aa}
\bibliography{biblio.bib}

\begin{appendix}

\setcounter{section}{1}
\begin{table}[t!]
\caption{\mbox{Comparison of literature data on atmospheric parameters of HD~37356.}}
\vspace{-0.6cm}
\label{tab:comparison}
{\small
\begin{center}    
\begin{tabular}{lllllll}        
\hline\hline
$T_\mathrm{eff}$\,(K) & $\log g$\,(cgs) & $\xi$\,(km\,s$^{-1}$) & $\varv \sin i$\,(km\,s$^{-1}$) & [O/H]$_\mathrm{CAS}$ & [Fe/H]$_\mathrm{CAS}$ & Reference\\
\hline
23300$\pm$800 & 3.89$\pm$0.07 & 5   & 28$\pm$6 & $-$0.24$\pm$0.08 & $-$0.42$\pm$0.32\tablefootmark{a} & \citet{Kilian92,Kilian94}\\
22370         & 4.13          & 9   & ...      & $-$0.09$\pm$0.05 & $-$0.22$\pm$0.09\tablefootmark{a} & \citet{CuLa94}\\
21400$\pm$630 & 3.78$\pm$0.16 & 1.1 & 17.5$\pm$3 & $-$0.14$\pm$0.06 & ...              & \citet{Lyubimkovetal13}\\
6940...15000\tablefootmark{b}     & 3.55...4.80 & ... & 29...307 & ...              & ...              & \citet{Kounkeletal19}\\
16858.2           & 4.218         & ... & ... & ...              & $-$2.500\tablefootmark{c}      & \citet{Abdurroufetal22}\\
17300.9           & 4.176             & ... & ... & ...              & $-$2.013\tablefootmark{c}      & \citet{Abdurroufetal22}\\
14486.8           & 3.5135            & ... & ... & ...              & +0.6058\tablefootmark{c}           & \citet{Gaia23}\\
22000$\pm$1000& 4.00$\pm$0.25 & 4.50$\pm$0.10 & 18.00$\pm$0.10 & +0.01$\pm$0.12 & $-$0.09$\pm$0.19 & \citet{Unaletal25}\\
22039$\pm$131 & 3.96$\pm$0.02 & 1.6$\pm$0.8   & 18.6$\pm$0.6 & +0.03$\pm$0.08 & $-$0.45$\pm$0.06 & this work\\  
\hline
\end{tabular}
\end{center}}
\vspace{-0.5cm}
\tablefoot{
\mbox{
\tablefoottext{a}{LTE values}
\tablefoottext{b}{range from analyses of 14 individual spectra}
\tablefoottext{c}{values relative to the solar iron abundance.}}}
\end{table}
\setcounter{section}{0}
\section{Comparison with literature data}\label{appendixA}
As one of the bright early-type stars in the Orion association, HD~37356 has been subject to a number of quantitative studies. Table~\ref{tab:comparison} summarises the atmospheric parameters derived, $T_\mathrm{eff}$, $\log g$, $\xi$ and $\varv \sin i$, and gives oxygen and iron abundances relative to CAS values, [X/H]$_\mathrm{CAS}$\,=\,$\log$(X/H)\,$-$\,$\log$(X/H)$_\mathrm{CAS}$, as examples of a volatile and a refractory element, respectively; references are also indicated. The studies split into two classes, targeted detailed analyses of individual early B-type stars, and bulk pipeline-determined data as part of surveys that investigated more than ten thousand stars, such as a multiplicity study within APOGEE-2 \citep{Kounkeletal19}, hundreds of thousands of stars, such as the full \mbox{APOGEE-2} release \citep{Abdurroufetal22}, or even more than a billion stars, such as \textit{Gaia} DR3 \citep{Gaia23}. The pipeline-derived parameters are not sensible for an early B-type star, more fitting to late or mid-B-ype stars, with the study of \citet{Kounkeletal19} analysing 14 individual high-quality spectra of HD~37356, finding even a range of $T_\mathrm{eff}$ values suitable to F- to mid-B-type stars, and some $\log g$ values locating the star even below the main sequence (which lies around a value of 4.3 at that $T_\mathrm{eff}$), and not to mention extremely high rotational velocities. The derived metallicities are in part far off the expectations for a young massive star, implying extreme halo population metallicity, or in the case of  the 4$\times$solar metallicity one may argue for a Bp signature. These pipeline-derived parameters may be statistically useful for the main targets of these surveys, such as solar-type stars and cool giants, but they have to be viewed with caution at present for OB-type stars, as apparently their modelling is subject to massive systematic errors \citep[for similar findings see][]{Aschenbrenneretal23,Wessmayeretal23}.

The four literature studies that targeted early B-type stars on the other hand roughly agree on $T_\mathrm{eff}$, $\log g$ and $\varv \sin i$ amongst each other and with our values. Our star HD~37356 is one of the cases where the use of photometric $T_\mathrm{eff}$ indicators \citep{CuLa94,Lyubimkovetal13} provides results similar to spectroscopic techniques. Modern microturbulence values of OB-type stars based on a range of indicators are systematically (much) smaller than older determinations \citep[see also][]{NiPr12,Aschenbrenneretal23,Wessmayeretal22,Wessmayeretal23}, and apparently also smaller than the determination based on a single indicator, in this case the \ion{Si}{iii} triplet \citep{Unaletal25}. The derived elemental abundances vary somewhat from each other, with \citet{Kilian92,Kilian94}, \citet{CuLa94} and \citet{Lyubimkovetal13} indicating overall sub-CAS (and therefore also sub-solar) values, while \citet{Unaletal25} reproduce CAS values. Notable with respect to the latter work is that they also determine atmospheric parameters for $\gamma$~Peg,~\,finding an approximately~\,20\%~\,higher~\,effective\\
\rule{0cm}{5.4cm}\\
temperature despite the similarity of the spectra stressed here. Unfortunately, no comparison of abundances can be made, as they refrain from a chemical analysis for $\gamma$~Peg.

In a situation when there are somewhat conflicting results available for a star, one has to resort to the basic principle in physics that a theory (in this case a model) that reproduces the experiment (here the observations) better has to be preferred over its competitors. Not only the pure empirical similarity of the spectra of HD~37356 and $\gamma$~Peg, except for the weaker spectral lines of the refractory elements, speaks in our favour. There is also a tight match of our model with all observational constraints: the detailed agreement of the synthetic spectrum with the observed one (see Appendix~\ref{appendixB}), including the series of hydrogen lines and multiple ionisation equilibria (\ion{C}{ii/iii}, \ion{O}{i/ii}, \ion{Ne}{i/ii}, \ion{Al}{ii/iii}, \ion{Si}{ii/iii/iv}, \ion{S}{ii/iii}, \ion{Fe}{ii/iii}), the tight reproduction of the SED, and the pulsational behaviour in comparison to that of $\gamma$~Peg.

\section{Spectral fit}\label{appendixB}
A comparison between the observed FEROS spectrum of HD~37356 from 3900\,{\AA} to 5250\,{\AA} and the region around H$\alpha$ with the global best-fitting model is shown in Fig.~\ref{fig:specfits} In addition to stellar lines, numerous diffuse interstellar bands (DIBs) are present, and other interstellar lines such as the \ion{Ca}{ii} H and K lines. The quality of the fit at this very high S/N and low $\varv \sin i$ of the observed spectrum is overall excellent, which implies that the line list is mostly complete \citep[missing lines involve, e.g. levels of the \ion{C}{ii} quartet spin system above the first ionisation threshold that are not included in the model atom;][]{NiPr06,NiPr08}, that the oscillator strengths are mostly correct, a proper determination of atmospheric parameters, and accurate and precise determination of non-LTE level populations (i.e. reliable model atoms). The largest differences arise from the DIBs and the interstellar lines, which are not modelled. Inherent to the models are inaccuracies due to the selected oscillator strengths, which are mostly responsible for the line-to-line abundance scatter. Uncertainties of less than 10-20\% should be typical for the oscillator strengths of many of the metal lines investigated here, but some may reach 50\%, see the quality markers assigned in the data collection of the National Institute of Standards and Technology, NIST\footnote{\url{https://www.nist.gov/pml/atomic-spectra-database}}. For a similar plot for our comparison star $\gamma$~Peg, see Figs.~9a-e of \citet{NiPr12}; note, however, that at that time the spectrum synthesis was not as comprehensive as achieved in the meantime.

\begin{figure*}[ht]
\centering
\includegraphics[width=.945\linewidth]{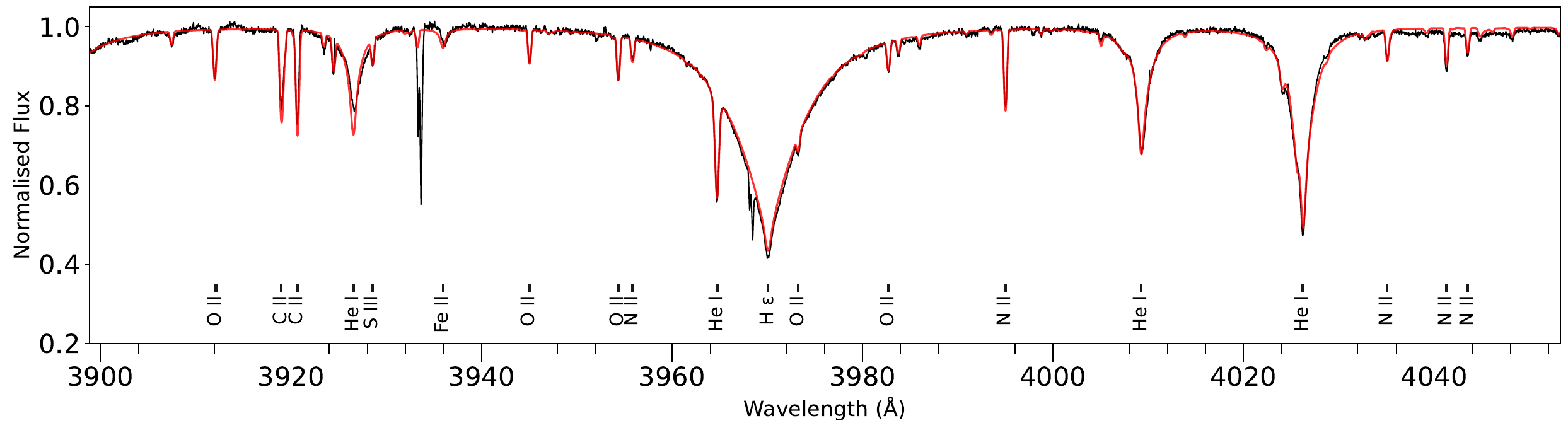}
\includegraphics[width=.945\linewidth]{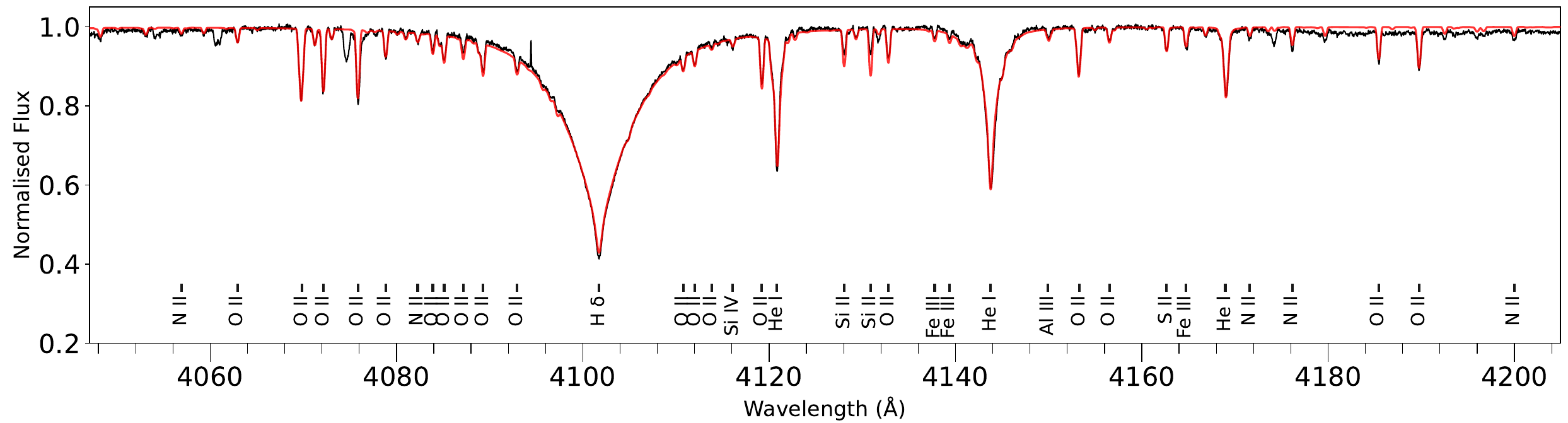}
\includegraphics[width=.945\linewidth]{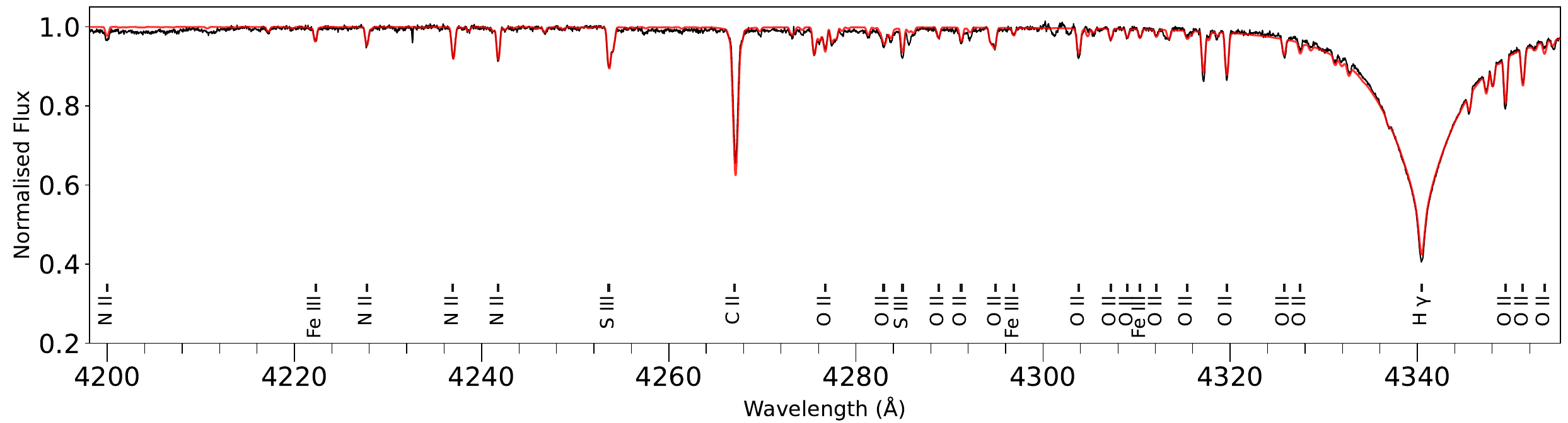}
\includegraphics[width=.945\linewidth]{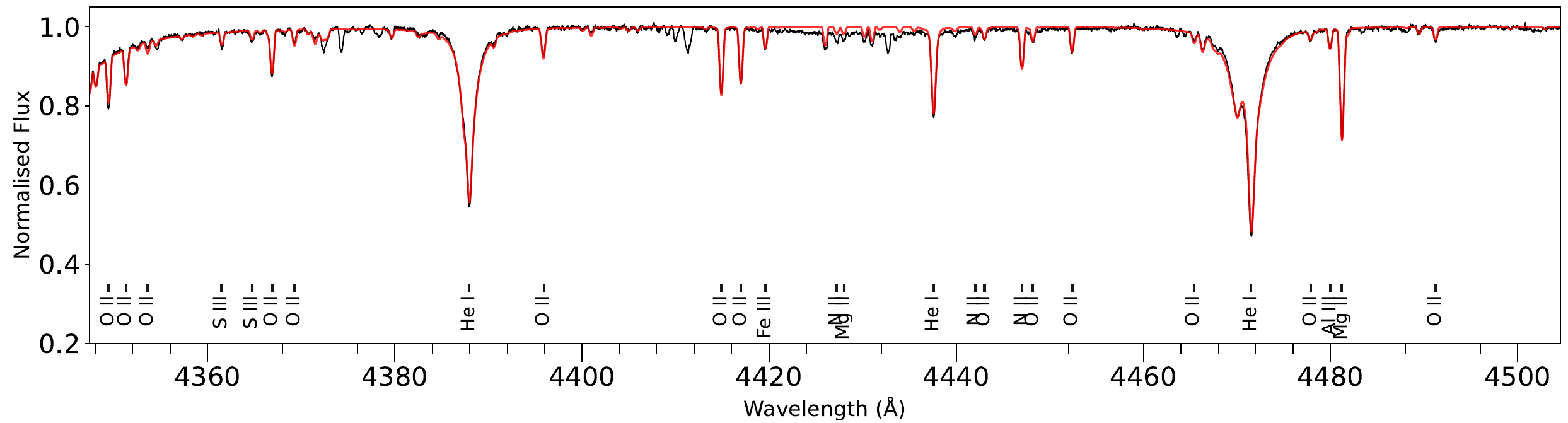}
\includegraphics[width=.945\linewidth]{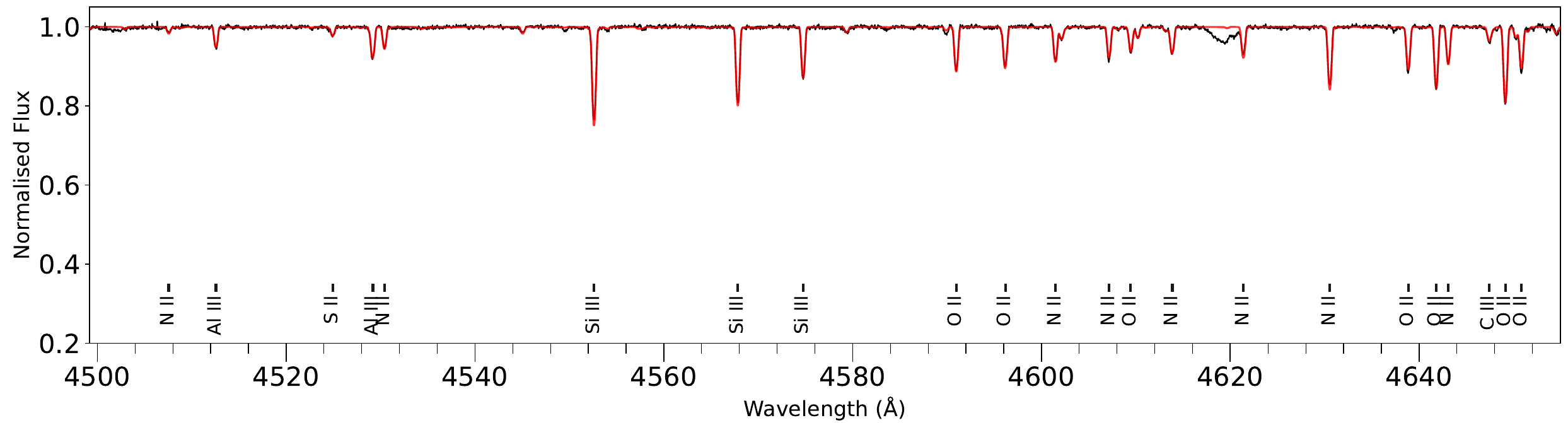}
\caption{Comparison between the observed spectrum of HD~37356 in black and the global best-fitting model in red. The strongest spectral lines are marked by solid lines. We note the presence of a few narrow interstellar lines and broader diffuse interstellar bands.}
\label{fig:specfits}
\end{figure*}
\begin{figure*}[ht]\ContinuedFloat
\centering
\includegraphics[width=.945\linewidth]{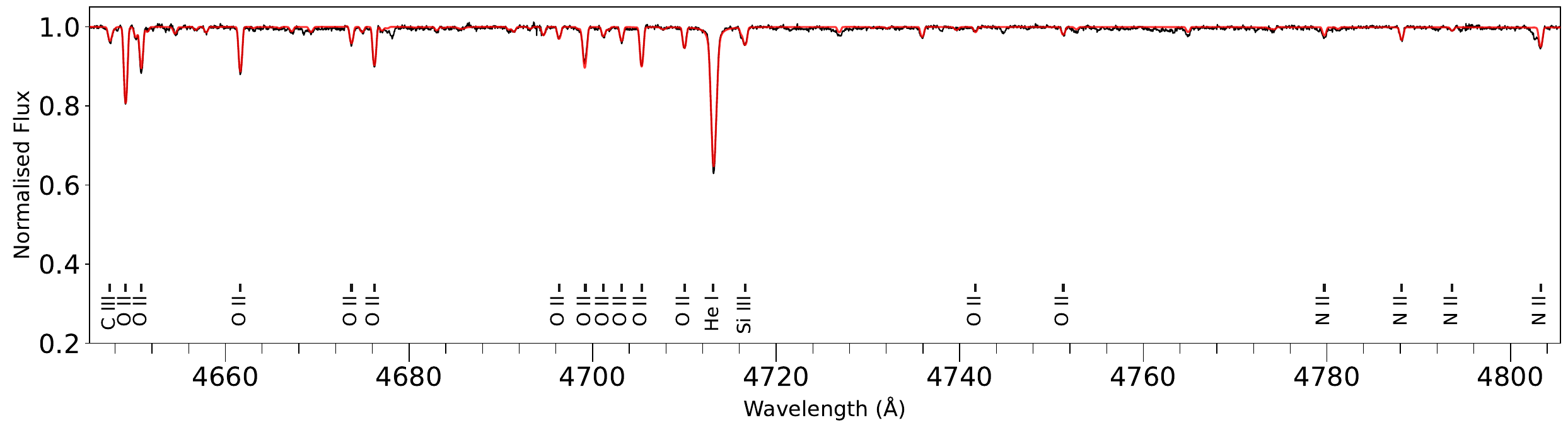}
\includegraphics[width=.945\linewidth]{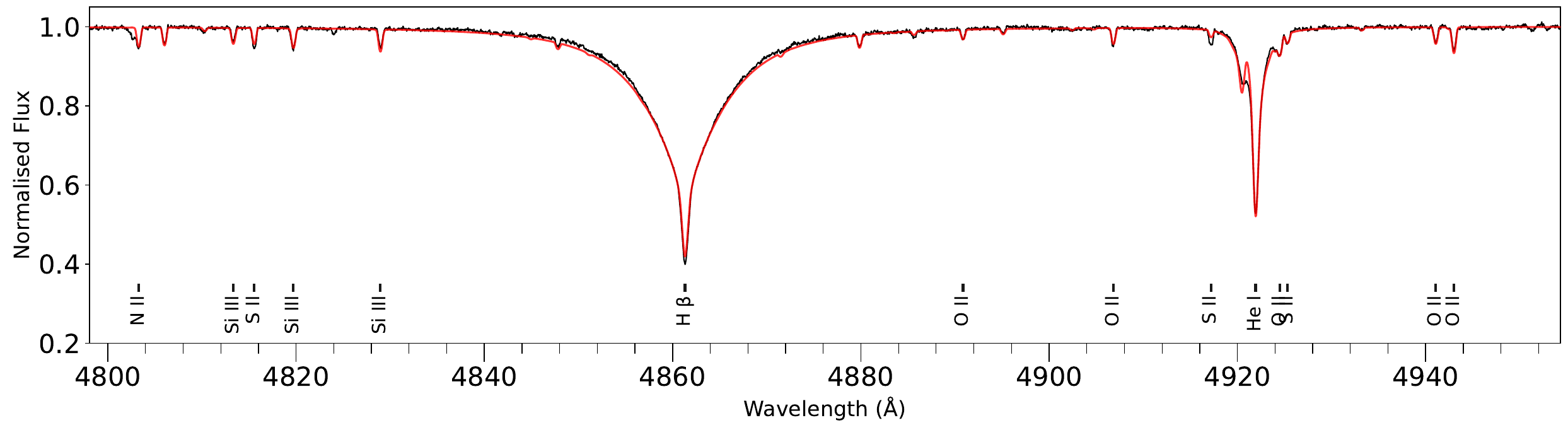}
\includegraphics[width=.945\linewidth]{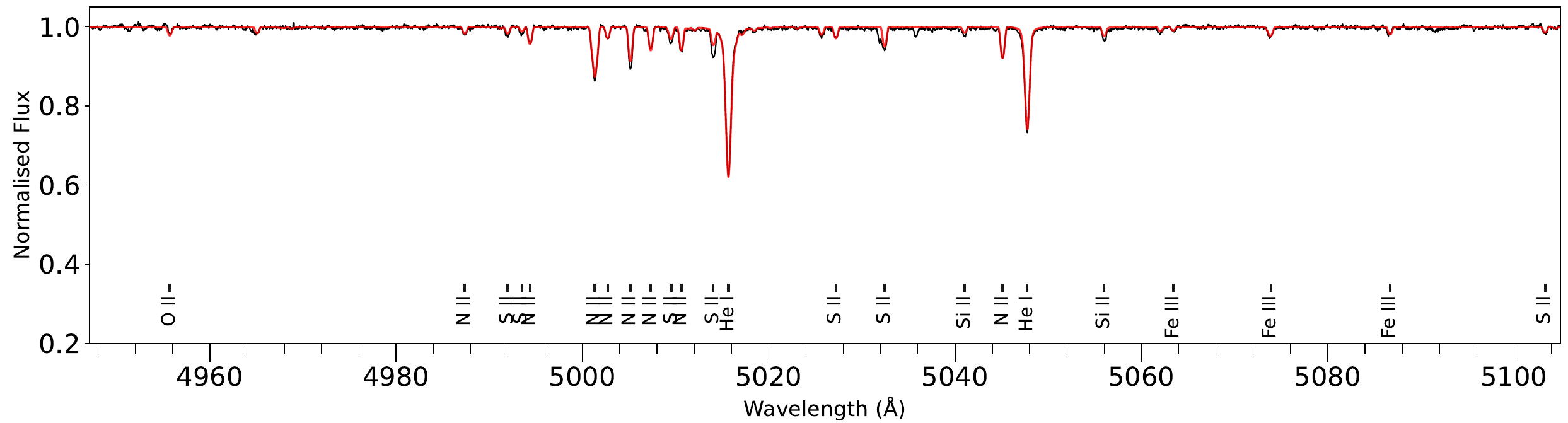}
\includegraphics[width=.945\linewidth]{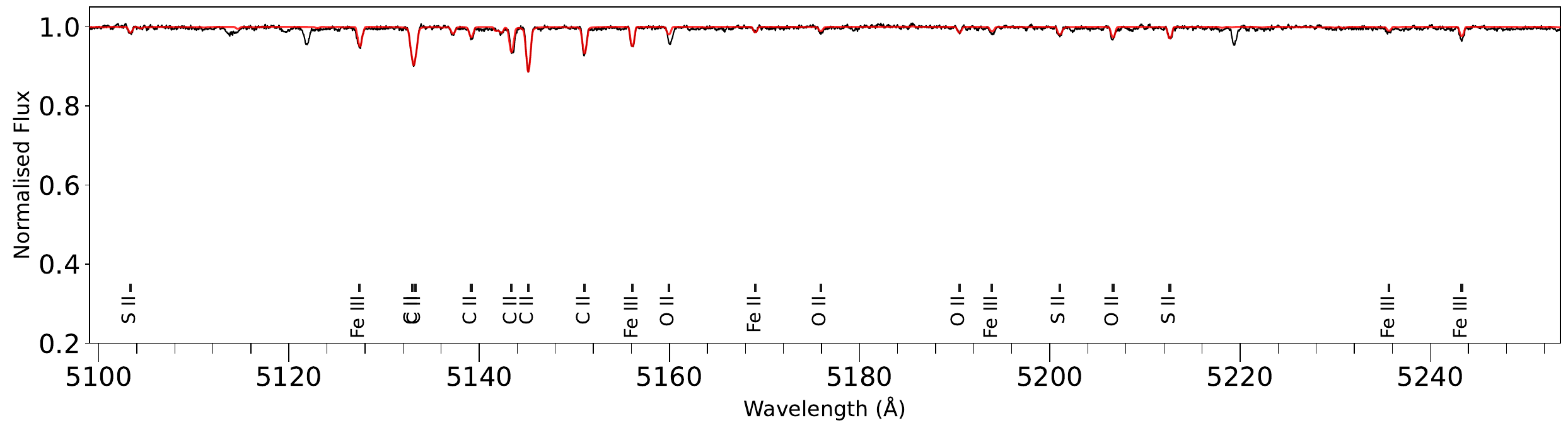}
\includegraphics[width=.945\linewidth]{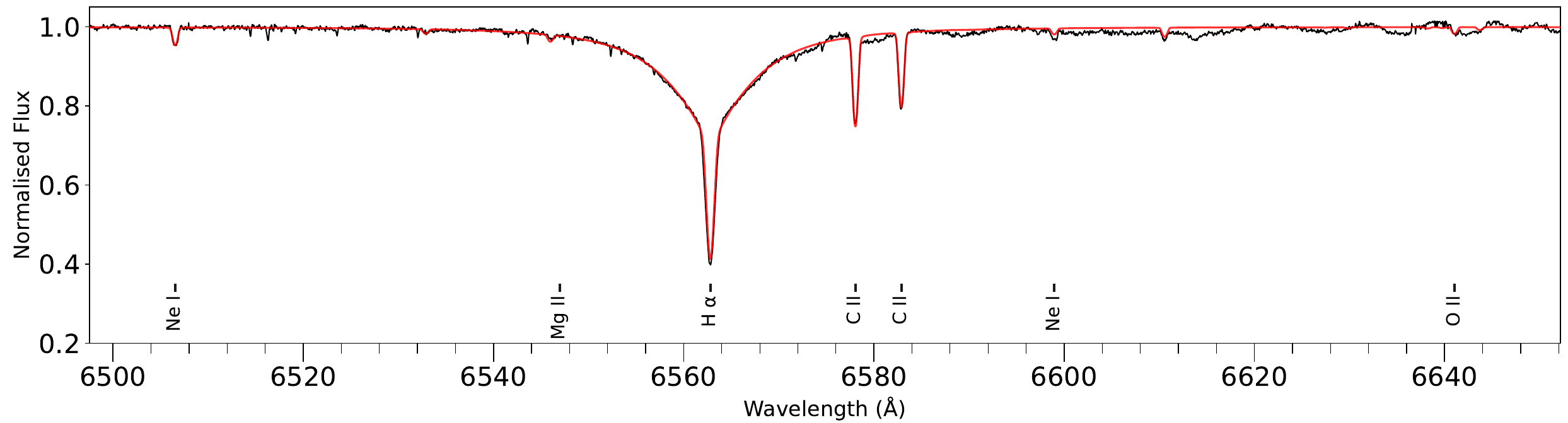}
\caption*{\raggedright Continued.}
\end{figure*}

\end{appendix}

\end{document}